\documentclass[twocolumn,epjc3]{svjour3}   
\pdfoutput=1

\smartqed  

\usepackage{amsmath}
\usepackage{graphicx}
\usepackage{bbold}
\usepackage{color}
\usepackage{subfigure}
\usepackage{multirow}
\usepackage{graphicx,bm}

\usepackage{dcolumn}
\usepackage{bm}
\usepackage{slashed}
\usepackage{epsfig}
\usepackage{hyperref}
\usepackage{verbatim}

\usepackage[utf8]{inputenc}
\usepackage[T1]{fontenc}

\newcommand{\beq}{\begin{eqnarray}}
\newcommand{\eeq}{\end{eqnarray}}

\newcommand{\bmp}{\noindent\begin{minipage}{16cm}}
\newcommand{\emp}{\end{minipage}\vskip 7mm} 

\newcommand{\be}{\begin{eqnarray}}
\newcommand{\ee}{\end{eqnarray}}

\newcommand{\SU}{\mbox{SU}}
\newcommand{\SO}{\mbox{SO}}
\newcommand{\SP}{\mbox{Sp}}
\newcommand{\UU}{\mbox{U}}

\begin{document}
\title{Towards a fundamental safe theory of composite Higgs and Dark Matter}

\author{Giacomo~Cacciapaglia\thanksref{e1,addr1}
        \and
       Teng~Ma\thanksref{e3,addr2,addr3}
                \and
       Shahram~Vatani\thanksref{e4,addr1}
                       \and
       Yongcheng~Wu\thanksref{e5,addr4}
}
\thankstext{e1}{g.cacciapaglia@ipnl.in2p3.fr}
\thankstext{e3}{t.ma@campus.technion.ac.il}
\thankstext{e4}{vatani@ipnl.in2p3.fr}
\thankstext{e5}{ycwu@physics.carleton.ca}

\institute{Universit\'{e} de  Lyon, Univ. Claude Bernard Lyon 1, CNRS/IN2P3, IP2I Lyon, UMR 5822, F-69622, Villeurbanne, France 
\label{addr1}
          \and
         CAS Key Laboratory of Theoretical Physics, Institute of Theoretical Physics, Chinese Academy of Sciences, Beijing 100190, China
\label{addr2}
        \and
        Physics Department, Technion -- Israel Institute of Technology, Haifa 3200003, Israel.
\label{addr3}
         \and
         Ottawa-Carleton Institute for Physics, Carleton University, 1125 Colonel By Drive, Ottawa, Ontario K1S 5B6, Canada.
\label{addr4}
}

\date{Received: date / Accepted: date}

\maketitle


\begin{abstract}
We present a novel paradigm that allows to define a composite theory at the electroweak scale that is well defined all the way up to any energy by means of safety in the UV. The theory flows from a complete UV fixed point to an IR fixed point for the strong dynamics (which gives the desired walking) before generating a mass gap at the TeV scale. We discuss two models featuring a composite Higgs, Dark Matter and partial compositeness for all SM fermions. The UV theories can also be embedded in a Pati-Salam partial unification, thus removing the instability generated by the $\UU(1)$ running. Finally, we find a Dark Matter candidate still allowed at masses of $260$~GeV, or $1.5 \sim 2$~TeV, where the latter mass range will be covered by next generation direct detection experiments. 
\end{abstract}

\maketitle

\section{Introduction}

The ultra-violet (UV) behaviour is crucial for a quantum field theory (QFT) to be predictive and fundamental
up to high scales~\cite{Wilson:1971bg,Wilson:1971dh}. The presence of fixed points in the renormalisation group
evolution of gauge and non-gauge couplings plays a central role in this. The prime example is Quantum
Chromo-Dynamics (QCD), which features a free fixed point where the gauge coupling vanishes in the UV~\cite{Gross:1973id,Politzer:1973fx}.
The possible existence of an interacting fixed point has been first proposed by S.~Weinberg in the context
of quantum gravity~\cite{Weinberg:1980gg}, but prematurely discarded for renormalisable QFTs. Until, F.~Sannino
and D.~Litim~\cite{Litim:2014uca} found a first example of perturbative interacting UV fixed point in a
theory with scalars and gauge-Yukawa couplings. Safe pure gauge theories with fermions have been obtained by employing
 resummation techniques for large number of fermionic flavours~\cite{Espriu:1982pb,PalanquesMestre:1983zy,Gracey:1996he}.  Recent progress can be
found in Refs~\cite{Antipin:2017ebo,Kowalska:2017pkt,Antipin:2018zdg,Alanne:2018ene,Alanne:2018csn}.
For a simple gauge group with $N_f$ fermions in the representation $r_f$, the resummed beta function
reads:
\beq \label{eq:RGEresummed}
\frac{\partial \ln K}{\partial \ln \mu} \equiv \beta (K) = \frac{2 K}{3} \left[ 1 + \sum_n \frac{1}{N_f^n} B^{(n)} \right] \,,
\eeq
where $K = N_f T(r_f) \alpha/\pi$. The presence of a UV fixed point is hinted by the fact that the first term
of the expansion $B^{(1)}$ has a negative pole at a finite value of $K$~\cite{Antipin:2017ebo}. Note that only
the first term in the large $N_f$ expansion is known.~\footnote{Ref.~\cite{Dondi:2020qfj} reported a first attempt to calculate the second term for abelian gauge symmetry.} The analysis based on resummation has been
recently challenged in Ref.~\cite{Alanne:2019vuk} but without conclusively disproving the existence of a fixed point~\cite{Sannino:2019vuf},
while studies on the lattice are still inconclusive~\cite{Leino:2019qwk}. In this work we will assume that the
presence of a pole in the large $N_f$ expansion is a sign of a genuine UV fixed point.

The least attractive feature of this class of asymptotically safe theories is the need for a
large multiplicity of fermion matter fields, as it can be seen in the attempts to build a safe extension of the
Standard Model~\cite{Mann:2017wzh,Pelaggi:2017abg}. Whilst this possibility is not ruled out experimentally, postulating the presence
of tens of new massive fermions at the multi-TeV scale for the sole purpose of changing the UV behaviour
of the theory contradicts the principle of minimality~\footnote{Quantum gravity effects may also be able
to drive the Standard Model interactions to a safe UV, see for instance Ref.~\cite{Eichhorn:2018yfc}.}. In this work we want to point
out a class of theories where the presence of large multiplicities of heavy fermions is required for another
crucial reason: the generation of masses for all Standard Model fermions in models of composite Higgs
with partial compositeness. In these models, minimality requires that there exists one composite operator
for each chiral fermionic field in the Standard Model. To generate such a spectrum of operators, an
underlying theory needs to contain a large number of preons, the fundamental degrees of freedom
that constitute the composite objects. Furthermore, considerations related to the hierarchy problem
lead to postulating fermionic preons.

A classification of underlying theories based on gauge-fermion interactions can be found in Ref.~\cite{Ferretti:2013kya}.
The main constraint on this model building effort is precisely the requirement that the theory shall remain
 confining at low energies~\cite{Ferretti:2016upr}. This limits the number of
underlying fermions to the ones responsible for giving mass to the top quark only. This problem is absent
in theories with scalar fields~\cite{Sannino:2016sfx} at the price of reintroducing the hierarchy problem
related to elementary scalar masses. In any case, these theories remain underlying descriptions of
the composite Higgs dynamics of top partial compositeness, but far from being true UV completions. In fact,
the origin of the light fermion masses as well as the source for the couplings generating the partial
compositeness remain absent. Our goal is therefore to take a decisive step towards addressing these
issues and being able to construct a genuine UV complete theory that can be trusted at arbitrarily high
energies, at least up to the Planck scale. In this perspective, providing a Dark Matter candidate
becomes a key ingredient.

In this work we present a new paradigm that allows to define composite Higgs models with underlying
fermions up to arbitrary high energies. The large number of fermions needed to give mass to all standard
quarks and leptons drives the theory to a complete UV interacting fixed point. The fermions
associated with the two light generations are supposed to have a large mass, thus explaining the lightness
of their partners compared to the electroweak scale. Once integrated out, the remaining degrees of freedom drive the
confining gauge interaction towards an Infra-Red (IR) fixed point~\cite{Banks:1981nn}. The resulting conformal window, similar in nature
to walking Technicolor~\cite{Holdom:1981rm,Yamawaki:1985zg}, allows to further split the scale of the heavy fermions where flavour effects also
arise, from the condensation and electroweak scales. The exit from the IR fixed point can be driven by
integrating out a subset of the remaining light fermions, leaving one of the models of Ref.~\cite{Ferretti:2013kya}
at low scale. In this framework, fundamental scalar fields can also be added in a natural way, as long as
their masses are close to the mass of the heaviest fermions~\cite{Pelaggi:2017wzr}, and they can be responsible
for generating the needed four-fermion interactions. The low-energy flavour mixing of the SM can therefore be traced
back to high-scale Yukawa couplings of scalars that, as we will show, are charged under the confining strong interactions. A Dark Matter
candidate can be easily included in this class of theories~\cite{Ma:2017vzm,Ballesteros:2017xeg,Balkin:2017aep}. The new scenario we discuss here,
therefore, allows to define composite models that can be as predictive as supersymmetric extensions
of the Standard Model.

This paper is organised as follows: after introducing the general set-up in Sec.~\ref{sec:model}, we describe in Sec.~\ref{sec:safe} how two
underlying models of top partial compositeness with Dark Matter can be extended to UV safety. In Sec.~\ref{scalars} we introduce the scalar
sector at high energy, responsible for generating the flavour couplings. In Sec.~\ref{PatiSalam} we embed the two models in a Pati-Salam
unification framework, thus eliminating the $\UU(1)$ problem. Finally, in Sec.~\ref{DM} we analyse the phenomenology of the Dark Matter
candidate, before offering our conclusions in Sec.~\ref{sec:conclusion}.

\section{Choosing the model}
\label{sec:model}

\begin{table*}[t]
\begin{center}
\begin{tabular}{c|c|c|c|c|c}
\hline\hline
$\psi$ irrep    & coset & pNGBs & pNGB EW charges &  models & $\mathcal{G}_{\rm HC}$ \\
\hline\hline
pseudo-real & $\SU(4)/\SP(4)$ & 5 & $2_{\pm 1/2} \oplus 1_0$  & M8-M9 & $\SP(4)$, $SO(11)$ \\\hline
real & $\SU(5)/\SO(5)$ & 14 & $2_{\pm 1/2} \oplus 3_{\pm1} \oplus 3_0 \oplus 1_0$ &  M1-M7 & $\SU(4)$, $\SP(4)$, $\SO(7)$, $\SO(9)$, $\SO(10)$  \\\hline
complex & $\SU(4)^2/\SU(4)$ & 15 & $2 \times 2_{\pm 1/2} \oplus 3_0 $ & M10-M12 & $\SO(10)$, $\SU(4)$, $\SU(5)$ \\
              &                                &      &     $\oplus 1_{\pm 1} \oplus 2 \times 1_0$ & MV & $\SU(3)$  \\     \hline
\end{tabular} \end{center}
\caption{Minimal cosets with a pNGB Higgs doublet arising from an underlying gauge-fermion theory. The fourth column shows the $\SU(2)_L$ {\it irrep}, with the hypercharge as subscript. The last three columns show some properties of the explicit models, with the nomenclature M1-M12 from Ref.~\cite{Belyaev:2016ftv}, and MV being the model from Ref.~\cite{Vecchi:2015fma}.} \label{tab:cosets}
\end{table*}

To define an underlying theory for composite Higgs models, we need to specify a confining hyper-colour (HC) gauge
symmetry, $\mathcal{G}_{\rm HC}$, and the irreducible representation ({\it irrep}) of the underlying fermions, $\psi_i$.
Furthermore, the electroweak (EW) quantum numbers of the $\psi_i$ should be suitably chosen such that a Higgs
doublet arises as a pseudo-Nambu Goldstone boson (pNGB) after confinement and chiral symmetry breaking. The partial
compositeness paradigm imposes a strong additional requirement: the presence of spin-1/2 bound states that
mix with the standard fermions.

Bound states of three $\psi$'s are  possible for the fundamental {\it irrep} of $\mathcal{G}_{\rm HC} = SU(3)$~\cite{Vecchi:2015fma},
as long as some of the fermions also carry QCD charges in addition to the EW ones. Another possibility, proposed in Ref.~\cite{Ferretti:2013kya},
is to sequester QCD charges to a second class of underlying fermions, $\chi_j$, transforming under a different $\mathcal{G}_{\rm HC}$
{\it irrep}. The benefit of this choice is that the breaking of the EW symmetry via vacuum
misalignment in the $\psi$-sector is decoupled from QCD, which shall not be broken. Furthermore, the spin-1/2
bound states that enter partial compositeness arise as chimera baryons~\cite{Ayyar:2017qdf} made of both
species of fermions, in the two alternative forms
\begin{equation}
\mathcal{B} = \langle \psi \psi \chi \rangle \quad \mbox{or} \quad \langle \psi \chi \chi \rangle\,.
\end{equation}
For each HC gauge group, the multiplicity of fermions, and their {\it irreps}, are limited by the requirement that the
theory remains asymptotically free, {\it i.e.} it confines at low energy, and outside the IR conformal
window~\cite{Dietrich:2006cm,Sannino:2009aw}, {\it i.e.} a mass gap is generated at low energy. Furthermore, the minimal number of $\psi$'s
is given by the requirement of having a Higgs doublet in the coset,
while the minimal $\chi$ sector needs to contain a QCD colour triplet and an anti-triplet in order to generate - at least -
the top mass. This leaves only 12 feasible models~\cite{Ferretti:2016upr} with minimal Higgs cosets, which we
denote M1 to M12, following Ref.~\cite{Belyaev:2016ftv}. The low energy models we consider here, and their key
features, are listed in Table~\ref{tab:cosets}.

To define a genuine UV completion for these models, the issue of Dark Matter cannot be avoided. The simplest possibility is
that one of the additional pNGBs may be stable. The minimal case is offered by the coset $\SU(4)_L \times SU(4)_R/\SU(4)$~\cite{Ma:2015gra,Ma:2017vzm}, which can be obtained in models M10-12, and the models of Refs~\cite{Vecchi:2015fma,Sannino:2016sfx}. As it was shown in Ref.~\cite{Ma:2015gra}, in the EW sector there is a unique $\mathbb{Z}_2$ parity that is conserved by the fermion condensate (if custodial symmetry is preserved) and by the EW gauging, as well as being anomaly free: it is defined in terms of charge conjugation in the $\psi$ sector plus a flavour rotation in the $\SU(4)$ flavour space. If the top couplings also respect this parity, the pNGB spectrum will contain several odd scalars, in particular a doublet and a triplet of $\SU(2)_L$ plus a neutral and a charged singlet. Such states mix, and the lightest neutral one plays the role of Dark Mater candidate (see Ref.~\cite{Ma:2015gra} for more details on the pNGB structure).

To extend the Dark $\mathbb{Z}_2$ parity in the case of partial compositeness, we need to make sure that the composite operator $\mathcal{B}$ that mixes with the top has well-defined transformation properties and contains even states with the same quantum numbers as the top quark fields.  As the Dark parity contains charge conjugation in the $\psi$-sector (but not $\chi$), it is crucial that the bound state contains two $\psi$'s: this simple fact rules out the case with HC-charged scalars of Ref.~\cite{Sannino:2016sfx}. Furthermore, the $\psi$-bilinear in $\mathcal{B}$ needs to be in a real {\it irrep} of $\mathcal{G}_{\rm HC}$ (ruling out M12 and the model of Ref.~\cite{Vecchi:2015fma}).
We are therefore left with the models M10 and M11, based on $\SO(10)_{\rm HC}$ and $\SO(6)_{\rm HC} \equiv \SU(4)_{\rm HC}$ respectively, with $\psi_i$ in the spinorial ($\bf Sp$) {\it irrep} and $\chi_j$ in the fundamental ($\bf F$).
For the top partners, there remains two choices that preserve the Dark parity: either a bi-fundamental of $\SU(4)_L\times \SU(4)_R$, which decomposes into a symmetric and an anti-symmetric of the unbroken $\SU(4)$, or a pair of symmetric {\it irreps}. As we will see in Section~\ref{DM}, only the first case leads to a feasible low energy model.

\section{A fundamental theory with UV safety and Dark Matter}
\label{sec:safe}

\begin{figure*}[tb]
\centering
\includegraphics[width=0.9\textwidth]{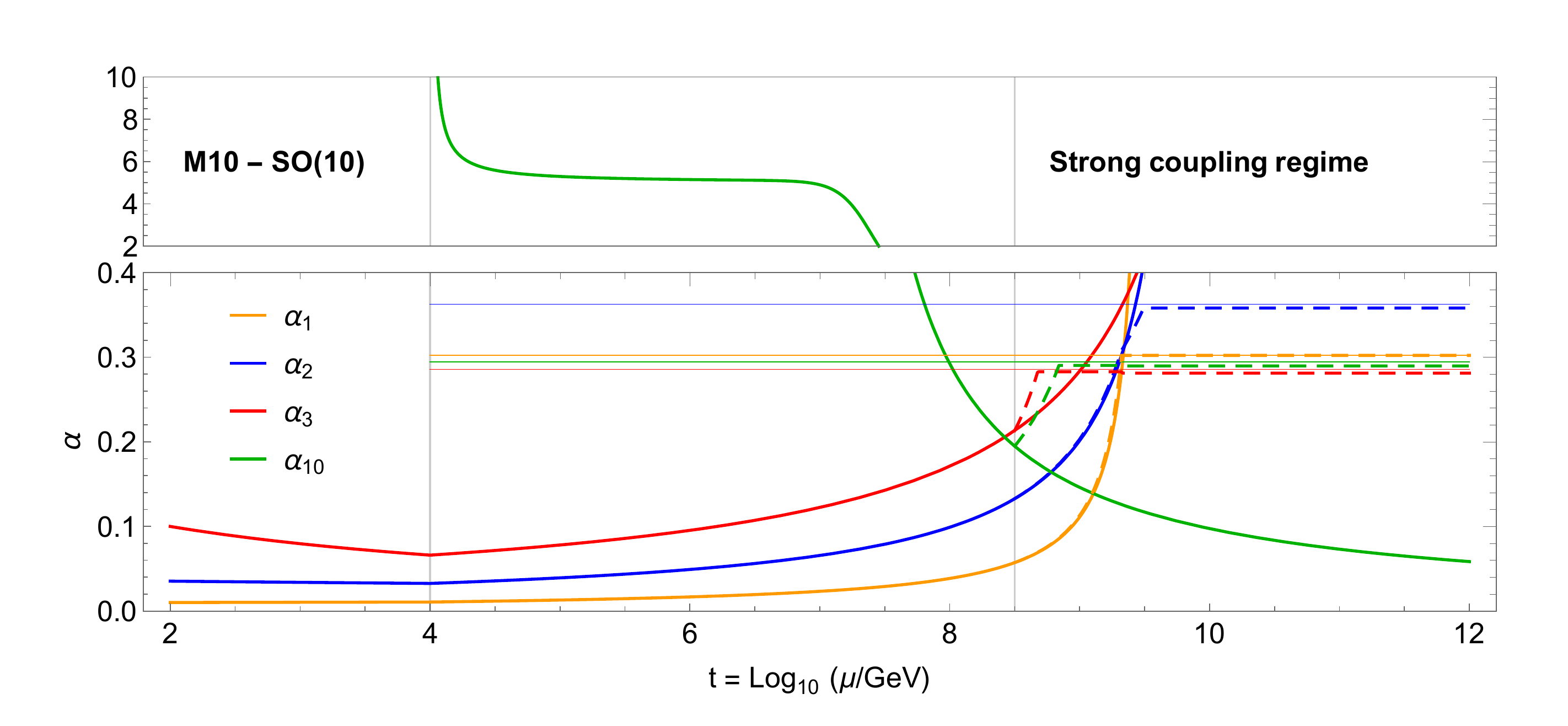}
\caption{Renormalisation group running of the gauge couplings $\alpha_i$ for model M10, $\mathcal{G}_{\rm HC} = \SO(10)$. The dashed lines show the effect of the large-$N_f$ resummation above $\Lambda_{\rm Fl} = 10^{8.5}$~GeV. The upper panel shows a cartoon of the running at strong coupling. }
\label{fig:running}
\end{figure*}

{ \color{red}
\begin{table}[htb]
\begin{center}
\begin{tabular}{c|c|c|c|c|c}
\hline\hline
    & $\SO(\mathcal{N})_{\rm HC}$ & $\SU(3)_c$ & $\SU(2)_L $ & $\UU(1)_Y$ & mass \\
\hline\hline
$\psi_Q$ & $\bf Sp$ & $1$ & $2$ & $0$ & \multirow{4}{*}{$\sim 0$}\\
$\psi_U$ & $\bf Sp$ & $1$ & $1$ & $1/2$ &  \\
$\psi_D$ & $\bf Sp$ & $1$ & $1$ & $-1/2$ &  \\
$\chi_u^3$ & $\bf F$ & $3$ & $1$ & $2/3$ & \\ \hline
$\chi_d^3$ & \multirow{2}{*}{$\bf F$} & $3$ & $1$ & $-1/3$ & \multirow{2}{*}{$\sim  \Lambda_{\rm HC}$}\\
$\chi_l^3$ &  & $1$ & $2$ & $-1/2$ & \\ \hline
$\chi_u^{1,2}$ & \multirow{3}{*}{$\bf F$} & $3$ & $1$ & $2/3$ & \multirow{3}{*}{$\begin{array}{c} \Lambda_{\rm Fl} \\ (\gg \Lambda_{\rm HC}) \end{array}$}\\
$\chi_d^{1,2}$ &  & $3$ & $1$ & $-1/3$ & \\
$\chi_l^{1,2}$ &  & $1$ & $2$ & $-1/2$ & \\ \hline
\end{tabular} \end{center}
\caption{Fermion content of the extended M10 ($\mathcal{N}=10$) and M11 ($\mathcal{N}=6$) - all fermions are Dirac spinors. } \label{tab:M10}
\end{table}
}
In the following, we will focus on M10 and M11: they only differ in the HC group, $\SO(10)_{\rm HC}$ versus $\SO(6)_{\rm HC}$.
The low-energy fermion content~\cite{Ferretti:2013kya} consists of 4 EW-charged $\psi$'s and a QCD-coloured triplet of $\chi$'s, as shown in the upper block
of Table~\ref{tab:M10}. These fermions characterise the composite states below the condensation scale $\Lambda_{\rm HC}$,
including the Higgs, the Dark Matter candidate and the top partners.

To complete the model, we will extend it by adding one appropriate $\chi$ for each standard fermion that acquires
mass via the Higgs mechanism, as shown in the remaining two blocks of Table~\ref{tab:M10}. We add the partners for the bottom quark and tau lepton
at a scale close to $\Lambda_{\rm HC}$, {\it i.e.} 5 additional $\chi$-flavours. This is enough to push the theory into the
conformal window (more details in~\ref{app:confo}): right above the condensation scale, therefore, the strong sector flows into a conformal phase where
the gauge coupling remains strong and slowly walking. This phase may ensure that the operators that mix to the light
generations acquire a largish anomalous dimension, allowing to sufficiently decouple the scale where they are introduced.
The $\chi$ fermions associated to the light generations are, in fact, introduced at a scale $\Lambda_{\rm Fl} \gg \Lambda_{\rm HC}$,
where flavour effects are also generated. 
Above $\Lambda_{\rm Fl}$, the number of fermions is such that the running of all
gauge couplings are not asymptotically free any more. 
The lepton partners, $\chi_l^{i}$, are chosen to be doublets of $\SU(2)_L$ for two reasons: on the one hand, their presence will assure that the $\SU(2)_L$ gauge coupling also runs into safety; on the other hand, the quantum numbers are such that chimera baryons containing $\chi_l^{i}$ also feature a neutral singlet, i.e. right-handed neutrinos, thus allowing to generate neutrino masses.
We will first study how the gauge couplings of these theories may flow to a UV safe fixed point.

Our set up differs from the ones considered in the literature (see Ref. \cite{Antipin:2017ebo}) in the fact that we have different
sets of fermions participating to the running of the four gauge couplings. Furthermore, for the $\SO(\mathcal{N})$ group, there are
two different \emph{irreps} that need to be taken into account~\cite{Vatani:2020}. For these reasons, we define the large-$N_f$ gauge couplings as follows:
\begin{equation}
K_i \equiv N_i \frac{\alpha_i}{\pi}\,, \;\; N_i = \sum_f n_f T (r_f)\,;
\end{equation}
with $i = 1,2,3,\mathcal{N}$ labelling the four gauge groups
(for $\UU(1)$, replace $T(r_f) \to Y_f^2$).
As there are many fermions in different {\it irreps} of the gauge groups, in our case we cannot define a unique $N_f$ valid for all gauge coupling running, but rather we need to define a different multiplicity $N_i$ for each group. 
We will assume that formally they are all of the same order. For the extended models in Table~\ref{tab:M10}, we find the multiplicity factors listed in Table~\ref{tab:Nis}.
For the running of the $\SO(\mathcal{N})$ gauge coupling, as there are two different \emph{irreps} that contribute, we follow the results in Ref.~\cite{Vatani:2020}:
%
\begin{multline} \label{eq:Bi}
\frac{B^{(1)}_i}{N_f} =  \frac{C_2 (G_i)}{N_i} \left( - \frac{11}{4} + G_1 (K_i) \right) + \\ \sum_{j=1}^{\mathcal{N}} \frac{c_{i,j}}{N_j} F_1 (K_j)\,,
\end{multline}
where the functions $F_1$ and $G_1$ are defined in~\ref{app:resum}, and $C_2 (G_i)$ is the Casimir of the adjoint of the $i$--th gauge group (for the abelian case, $C_2 (G_1) = 0$). Note that the $-11/4$ term corresponds to the one loop contribution of gauge bosons.
The coefficients $c_{i,j}$ are all positive, and their values can be found in~\ref{app:resum}.

\begin{table}[htb]
\begin{center}
\begin{tabular}{c|c|c|c}
\hline\hline
    & $\SO(\mathcal{N})_{\rm HC}$ & M10 & M11 \\
\hline\hline
$N_{\mathcal{N}}$ & $24 + 2^{\frac{\mathcal{N}-4}{2}}$ & $32$ & $26$ \\
$N_3$ & $3 (\mathcal{N}+1)$ & $33$ & $21$ \\
$N_2$ & $3 + \frac{3}{2} \mathcal{N} + 2^{\frac{\mathcal{N}-4}{2}}$ & $26$ & $14$ \\
$N_1$ & $5 + 6.5 \mathcal{N} + 2^{\frac{\mathcal{N}-4}{2}}$ & $78$ & $46$ \\ \hline 
\end{tabular} \end{center}
\caption{Multiplicity factors for the resummation of the four gauge couplings above $\Lambda_{\rm Fl}$. The numerical values in the last two columns refer to M10 ($\mathcal{N}=10$) and M11 ($\mathcal{N}=6$).} \label{tab:Nis}
\end{table}

A key property of the above result is that the
function $G_1 (K)$, relevant for non-abelian gauge couplings, has a pole at negative values for $K=3$, while $F_1 (K)$ has a negative pole at $K_1=15/2$, while the resummation
fails for coupling values above the pole. This feature, thus, acts as a barrier for the evolution of the respective coupling
towards the UV, hinting at the presence of an interacting UV fixed point~\cite{Antipin:2017ebo}. In other words, if the value of the coupling
at the threshold $\Lambda_{\rm Fl}$ is below the pole, the evolution towards the UV will stop at that value
where the beta function vanishes and the theory approaches a fixed point.
We, therefore, expect the UV fixed point to arise at $K_i = 3$ for non-abelian groups, and $K_1 = 15/2$ for the abelian one.
The condition for the model to have a UV safe fixed point for all gauge couplings is that their value is below
the pole, i.e.
\begin{equation}
\alpha_i < \frac{3 \pi}{N_i}\;\; \mbox{for}\;\; i\neq 1\,,\quad  \mbox{and}\;\; \alpha_1 < \frac{15\pi}{2N_1}\,.
\end{equation}
The above conditions provide an upper bound on $\Lambda_{\rm Fl}$ due to the fact that some
of the gauge couplings increase towards the UV above $\Lambda_{\rm HC}$. On the other hand, 
an indirect lower bound derives from flavour physics, which gives
$\Lambda_{\rm Fl} > 10^{5}$~TeV for generic flavour violating effects.

In Fig.~\ref{fig:running} we show the running of the 4 gauge couplings above the EW scale for the model M10, assuming $\Lambda_{\rm HC} = 10$~TeV (solid lines). While the $\SO(10)$ gauge coupling $\alpha_{10}$ is asymptotically free, the other two run into a Landau pole below $10^{10}$~GeV.  Furthermore, it is the QCD coupling $\alpha_3$ that crosses the UV-safe threshold first, thus setting the maximum allowed value of $\Lambda_{\rm Fl}$ right below $10^9$~GeV.
In the numerical example, we added the complete set of fermions at a scale $\Lambda_{\rm Fl} = 10^{8.5}$~GeV: the modified running is plotted in dashed lines, clearly showing how the gauge couplings approach the UV fixed values. Note that they are a bit below the predicted ones: this is due to the backreaction of the $\UU(1)$ pole on the running of the non-abelian couplings, due to the fact that $F_1 (K_1)$ in Eq.~\eqref{eq:Bi} also has a pole at the fixed point.
In the upper panel we illustrate the running of $\alpha_{10}$ in the strong coupling regime, which features a walking region between $10^4$ and $10^7$~GeV. This part of the plot, being non-perturbative, can only be confirmed by lattice calculations along the lines of Refs~\cite{Ayyar:2017qdf,Bennett:2017kga,Lee:2018ztv,Witzel:2018gxm}.
These results show that the model M10 allows for a narrow mass window where the fermions at $\Lambda_{\rm Fl}$ can be added, squeezed between the flavour bounds and the limit from UV safety.

A similar analysis can be done for the model M11, based on $\mathcal{G}_{\rm HC} = \SO(6)$: in this case, it is the $\UU(1)$ coupling $\alpha_1$ that crosses the threshold first at a scale around $10^{13}$~GeV, while the $\alpha_2$ and $\alpha_3$ run much slower. This model can therefore allow for a larger flavour scale, and a wider walking window for $\alpha_6$.

The models we have studied here, however, are not truly UV complete, because the dynamics generating the partial compositeness four-fermion interactions is not included. In the following sections we will discuss how scalar mediators can do the job.

\section{High-scale scalar mediation, and the $\UU(1)$ problem.}
\label{scalars}

\begin{table}[htb]
\begin{center}
\begin{tabular}{c|c|c|c|c|c}
\hline\hline
    & $\SO(\mathcal{N})_{\rm HC}$ & $\SU(3)_c$ & $\SU(2)_L $ & $\UU(1)_Y$ & mass \\
\hline\hline
$\phi_{q}^a$ & $\bf Sp$ & \multirow{2}{*}{$3$} & \multirow{2}{*}{$1$} & \multirow{2}{*}{$1/6$} & \multirow{2}{*}{$\sim \Lambda_{\rm Fl}$}\\
$\bar{\phi}_{q}^a$ & $\bf \overline{Sp}$ & & &  &  \\
\hline
$\phi_{l}^a$ & $\bf Sp$ & \multirow{2}{*}{$1$} & \multirow{2}{*}{$1$} & \multirow{2}{*}{$-1/2$} & \multirow{2}{*}{$\sim \Lambda_{\rm Fl}$}\\
$\bar{\phi}_{l}^a$ & $\bf \overline{Sp}$ & & &  &  \\
\hline
\end{tabular} \end{center}
\caption{Scalar mediators for lepton and quark partial compositeness.} \label{tab:scalT}
\end{table}

The four-fermion interactions responsible for the partial compositeness couplings at low energy can be generated
via scalar mediation. This is acceptable in this class of models because scalar masses are ``natural'' if they are close
to the largest fermion mass in the model \cite{Pelaggi:2017wzr}, namely $m_\phi \approx \Lambda_{\rm Fl}$. The only additional condition
would be to check that all new couplings in the scalar sector, i.e. Yukawas and quartic couplings, also run to a UV
safe fixed point. In this and in the next sections we will address this question.

Firstly, in order to preserve the Dark $\mathbb{Z}_2$, it is necessary to add pairs of scalar fields that have the same 
quantum numbers under the SM gauge symmetries, while they are in conjugate $\bf Sp$ and $\bf \overline{Sp}$ \emph{irreps}
of the strong $\SO(\mathcal{N})$. One minimal set of mediators is shown in Table~\ref{tab:scalT}, where $a=1,2,3$ is an index
running over the SM generations. These four fields allow to add the following Yukawa couplings above $\Lambda_{\rm Fl}$:
\begin{eqnarray}
\mathcal{L}_{\rm Yuk, q} &=& \left\{ \lambda_q^{ab}\ \phi_q^{b,\ast} [q_L^a]_l [\psi_Q]_l + \lambda_u^{ab} \phi_q^{b,\ast} [u_R^a]_r [\psi_D]_r + \right. \nonumber \\
& \phantom{\frac{1}{2}}& \phantom{xxxxxxxx} \lambda_d^{ab} \phi_q^{b,\ast} [d_R^a]_r [\psi_U]_r + \nonumber \\
& \phantom{\frac{1}{2}}& \xi_{ur}^{ab}\ \phi_q^a [\chi_u^b]_l^c [\psi_U]_r + \xi_{ul}^{ab}\ \phi_q^a [\chi_u^b]_r^c [\psi_U]_l + \nonumber \\
& \phantom{\frac{1}{2}}& \xi_{dr}^{ab}\ \phi_q^a [\chi_d^b]_l^c [\psi_D]_r + \xi_{dl}^{ab}\ \phi_q^a [\chi_d^b]_r^c [\psi_D]_l + \nonumber \\
& \phantom{\frac{1}{2}}& \bar{\lambda}_q^{ab}\ \overline{\phi}_q^{b,\ast} [q_L^a]_l [\psi_Q]_r^c + \bar{\lambda}_u^{ab} \overline{\phi}_q^{b,\ast} [u_R^a]_r [\psi_U]_l^c + \nonumber \\
& \phantom{\frac{1}{2}}& \phantom{xxxxxxxx} \bar{\lambda}_d^{ab} \overline{\phi}_q^{b,\ast} [d_R^a]_r [\psi_D]_l^c + \nonumber \\
&  \phantom{\frac{1}{2}}& \bar{\xi}_{ur}^{ab}\ \overline{\phi}_q^a [\chi_u^b]_l^c [\psi_D]_l^c + \bar{\xi}_{ul}^{ab}\ \overline{\phi}_q^a [\chi_u^b]_r^c [\psi_D]_r^c + \nonumber \\
&  \phantom{\frac{1}{2}}& \left. \bar{\xi}_{dr}^{ab}\ \overline{\phi}_q^a [\chi_d^b]_l^c [\psi_U]_l^c + \bar{\xi}_{dl}^{ab}\ \overline{\phi}_q^a [\chi_d^b]_r^c [\psi_U]_r^c \right\}\,,
\label{eq:Yukq}
\end{eqnarray}
and
\begin{eqnarray}
\mathcal{L}_{\rm Yuk, l} &=& \left\{ \lambda_l^{ab}\ \phi_l^{b,\ast} [l_L^a]_l [\psi_Q]_l + \lambda_e^{ab} \phi_l^{b,\ast} [e_R^a]_r [\psi_U]_r + \right. \nonumber \\
& \phantom{\frac{1}{2}}& \phantom{xxxxxxxx} \lambda_\nu^{ab} \phi_l^{b,\ast} [\nu_R^a]_r [\psi_D]_r + \nonumber \\
& \phantom{\frac{1}{2}}& \xi_{er}^{ab}\ \phi_l^a [\chi_l^b]_l^c [\psi_Q]_r + \xi_{el}^{ab}\ \phi_l^a [\chi_l^b]_r^c [\psi_Q]_l + \nonumber \\
& \phantom{\frac{1}{2}}& \bar{\lambda}_l^{ab}\ \overline{\phi}_l^{b,\ast} [l_L^a]_l [\psi_Q]_r^c + \bar{\lambda}_e^{ab} \overline{\phi}_l^{b,\ast} [e_R^a]_r [\psi_D]_l^c + \nonumber \\
& \phantom{\frac{1}{2}}& \phantom{xxxxxxxx} \bar{\lambda}_\nu^{ab} \overline{\phi}_l^{b,\ast} [\nu_R^a]_r [\psi_U]_l^c + \nonumber \\
&  \phantom{\frac{1}{2}}& \left. \bar{\xi}_{er}^{ab}\ \overline{\phi}_l^a [\chi_l^b]_l^c [\psi_Q]_l^c + \bar{\xi}_{el}^{ab}\ \overline{\phi}_l^a [\chi_l^b]_r^c [\psi_Q]_r^c  \right\}\,;
\label{eq:Yukl}
\end{eqnarray}
where a sum over the SM flavour indices is left understood, the subscripts $[.]_{l/r}$ indicate respectively the left and right-handed chiralities, and the superscript $[.]^c$ the charge-conjugation.
The Dark parity is preserved as long as $\lambda = \bar{\lambda}$ and $\xi = \bar{\xi}$: as we will see this condition is renormalisation evolution invariant, 
thus it is preserved at all scales once it is imposed at $\mu = \Lambda_{\rm Fl}$. We also need to impose that the two scalars have the same mass. 
Once they are integrated out, they generate appropriate four-fermion interactions for all SM fermions. As an example, for the top (up-type quarks), among others:
\begin{multline}
\mathcal{L}_{\Lambda_{\rm Fl}} = \frac{\kappa_q^{ab}}{\Lambda_{\rm Fl}^2} \Big( [q^a_L]_l \cdot [\psi_Q]_l\ [\chi_u^b]_l^c \cdot [\psi_U]_r +  \\
 \phantom{xxxxxxxxxxxxxxxxx}  [q_L^a]_l \cdot [\psi_Q]_r^c \ [\chi_u^b]_l^c \cdot [\psi_D]_l^c \Big) +\\
 \frac{\kappa_u^{ab}}{\Lambda_{\rm Fl}^2} \Big( [u^a_R]_r \cdot [\psi_D]_r\ [\chi_u^b]_r^c \cdot [\psi_U]_l +  \\
  [u_R^a]_r \cdot [\psi_U]_l^c \ [\chi_u^b]_r^c \cdot [\psi_D]_r^c \Big)\,,  \label{eq:4fermion}
\end{multline}
with
\begin{equation}
\frac{\kappa_q^{ab}}{\Lambda_{\rm Fl}^2} = \lambda_q^{ac} [m_\phi^2]^{-1}_{cd} \xi_q^{db}\,, \quad \frac{\kappa_u^{ab}}{\Lambda_{\rm Fl}^2} = \lambda_u^{ab} [m_\phi^2]^{-1}_{cd} \xi_u^{db}\,.
\end{equation}
The mediators and Yukawa couplings have been selected such that the four-fermion interactions in Eq.~\eqref{eq:4fermion} generate a coupling of the top fields to 
a composite baryon in the {\it irrep} $(\bf{4}, \bf{4}) \oplus (\bar{\bf{4}}, \bar{\bf{4}})$ of the global symmetry $\SU(4)_L \times SU(4)_R$, which we will use in Section~\ref{DM} for the Dark Matter study.

The contribution of the scalars above $\Lambda_{\rm Fl}$ will not affect significantly the running of the gauge couplings. The running of the Yukawa couplings above $\Lambda_{\rm Fl}$ follows the calculations done in Ref.~\cite{Antipin:2018zdg}: it has been observed that the dominant contribution is due to the $\UU(1)$ gauge coupling once it has approached its fixed point, as the contribution to the Yukawa beta function has a pole at the same position.
For all Yukawas $y_i$, the beta function can, therefore, be approximated by
\beq
\beta (y_i) \approx  - y_i  \frac{15}{16 \pi^2 N_1} (2 d_{y_i,1} + 15 d'_{y_i,1})  \frac{1}{\frac{15}{2} - K_1}\,,
\eeq
where $K_1$ is exponentially close (from below) to $15/2$, and
\beq
d_{y_i,1} = Y_\phi^2 + 2 Y_{f1} Y_{f2}\,, \quad d'_{y_i,1} = \frac{(Y_{f1}-Y_{f2})^2}{6}\,, 
\eeq
with $Y_x$ being the hypercharges of the scalar and fermions in the Yukawa coupling $y_i$. Thus, if
\beq
X_{y_i} = 2 d_{y_i,1} + 15 d'_{y_i,1} > 0\,,
\eeq
the Yukawa coupling $y_i$ runs to zero in the UV. In our model, we find the values in Table~\ref{tab:runU1}, which 
show that all the Yukawa couplings in Eqs~\eqref{eq:Yukq} and~\eqref{eq:Yukl} are asymptotically free.

\begin{table}[htb]
\begin{center}
\begin{tabular}{c|c|c|c}
\hline\hline
$y_i$    & $d_{y_i,1}$ & $d'_{y_i,1}$ & $X_{y_i}$  \\
\hline\hline
$\lambda_q,\ \bar{\lambda}_q$ & $1/36$ & $1/216$ & $1/16$ \\ \hline
$\begin{array}{c} \lambda_u,\ \xi_{ur},\ \xi_{ul}, \\
\bar{\lambda}_u,\ \bar{\xi}_{ur},\ \bar{\xi}_{ul} \end{array}$  & $-23/36$ & $49/216$ & $17/16$ \\ \hline
$\begin{array}{c} \lambda_d,\ \xi_{dr},\ \xi_{dl}, \\ 
\bar{\lambda}_d,\ \bar{\xi}_{dr},\ \bar{\xi}_{dl} \end{array}$ & $-11/36$ & $25/216$ & $9/16$ \\ \hline
$\lambda_l,\ \lambda_\nu,\ \bar{\lambda}_l,\ \bar{\lambda}_\nu$ & $1/4$ & $1/24$ & $9/16$ \\ \hline
$\begin{array}{c} \lambda_e,\ \xi_{er},\ \xi_{el}, \\ 
\bar{\lambda}_e,\ \bar{\xi}_{er},\ \bar{\xi}_{el} \end{array}$ & $-3/4$ & $3/8$ & $33/16$ \\
\hline
\end{tabular} \end{center}
\caption{Beta function coefficients for the Yukawas in Eqs~\eqref{eq:Yukq} and~\eqref{eq:Yukl}.} \label{tab:runU1}
\end{table}

While the $\UU(1)$ fixed point drives the Yukawas to be asymptotically free, it is well established that it has a dangerous effect on scalar quartic couplings, which are driven to a Landau pole~\cite{Antipin:2018zdg}. We can address this issue by partly unifying the $\UU(1)$ into a non abelian gauge group, and our model offers an elegant path via a Pati-Salam structure, as discussed in the next section.

\section{Pati-Salam UV-safe completion.}
\label{PatiSalam}

\begin{figure*}[tb]
\centering
\includegraphics[width=0.9\textwidth]{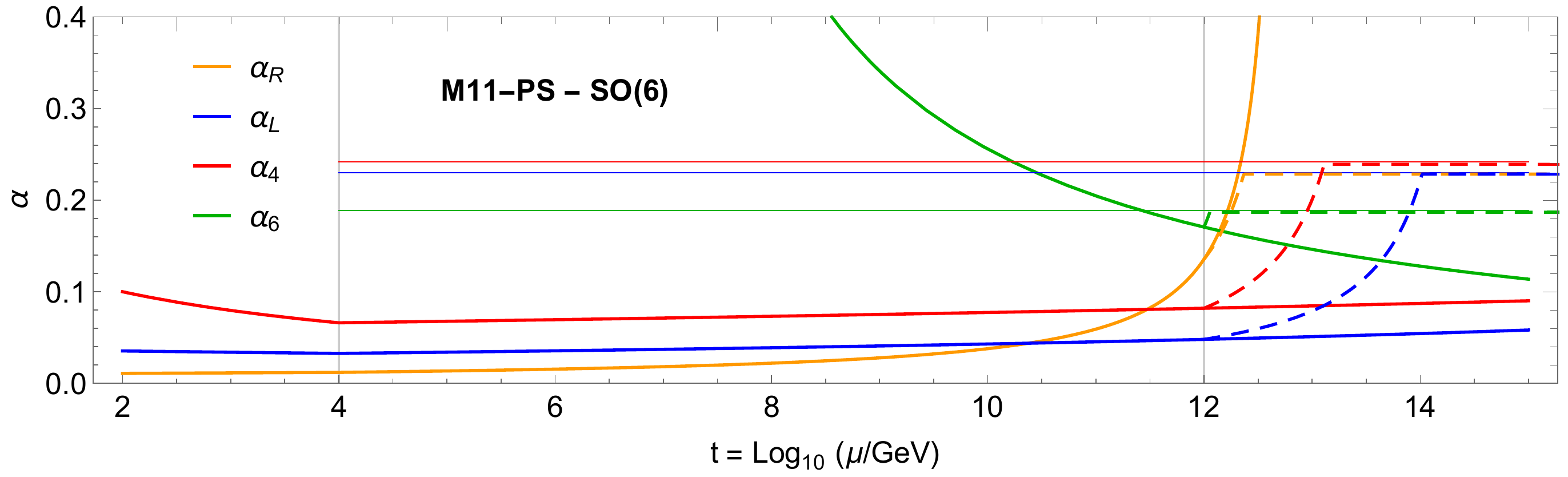}
\caption{Renormalisation group running of the gauge couplings $\alpha_i$ for model M11-PS, $\mathcal{G}_{\rm HC} = \SO(10)$. The dashed lines show the effect of the large-$N_f$ resummation above $\Lambda_{\rm Fl} = 10^{8.5}$~GeV. The upper panel shows a cartoon of the running at strong coupling. }
\label{fig:runningPS}
\end{figure*}

\begin{table}[htb]
\begin{center}
\begin{tabular}{c|c|c|c|c|c}
\hline\hline
    & $\SO(\mathcal{N})_{\rm HC}$ & $\SU(4)$ & $\SU(2)_L $ & $\SU(2)_R$ &  \\
\hline\hline
$\omega_L^a$ & $\bf 1$ & $4$ & $2$ & $1$ & $q_L^a,\ l_L^a$ \\
$\omega_R^a$ & $\bf 1$ & $4$ & $1$ & $2$ & $\begin{array}{c} u_R^a,\ d_R^a,\\ e_L^a,\ \nu_R^a \end{array}$ \\ \hline
$\Psi_{L}$ & $\bf Sp$ & $1$ & $2$ & $1$ & $\psi_Q$\\
$\Psi_R$ & $\bf Sp$ & $1$ & $1$ & $2$ & $\psi_U,\ \psi_D$ \\ \hline
$\Xi_R^a$ & $\bf F$ & $4$ & $1$ & $2$ & $\chi_u^a,\ \chi_d^a$\\
$\Xi_L^a$ & $\bf F$ & $4$ & $2$ & $1$ & $\chi_l^a$\\ \hline
$\Phi^a$ & $\bf Sp$ & $4$ & $1$ & $1$ & $\phi_q^a,\ \phi_l^a$ \\
$\overline{\Phi}^a$ & $\bf \overline{Sp}$ & $4$ & $1$ & $1$ & $\overline{\phi}_q^a,\ \overline{\phi}_l^a$ \\ \hline 
$\varphi_{\rm PS}$ & $\bf 1$ & $4$ & $1$ & $2$ & $-$ \\ \hline
\end{tabular} \end{center}
\caption{Fermion content of the Pati-Salam extended M10-PS ($\mathcal{N}=10$) and M11-PS ($\mathcal{N}=6$) - all fermions $\omega$, $\Psi$ and $\Xi$ are Dirac spinors.} \label{tab:MPS}
\end{table}

To remove the destabilising effect of the $\UU(1)$ pole on scalar quartic couplings, we can embed the model above $\Lambda_{\rm Fl}$ into a Pati-Salam~\cite{Pati:1974yy} 
partial unification for the SM interactions~\cite{Molinaro:2018kjz}. The new models, that we dub M10-PS and M11-PS, feature the field content in Table~\ref{tab:MPS}.
The Pati-Salam gauge group, $\SU(4) \times SU(2)_L \times \SU(2)_R$ is broken by a scalar field $\varphi_{\rm PS}$, with a vacuum expectation value of the order of $\Lambda_{\rm Fl}$.
To study the UV properties of this model, we can calculate the beta functions following the same procedure highlighted in Section~\ref{sec:safe}, with the important difference that the contribution of other gauge couplings remains negligible due to the absence of an abelian group.
The new $N_i$ are given in Table~\ref{tab:NisPS}: they are substantially larger than the corresponding ones in the previous models, indicating lower values for the UV fixed points. This is potentially dangerous, as the upper limit on $\Lambda_{\rm Fl}$ will tend to decrease.

\begin{table}[htb]
\begin{center}
\begin{tabular}{c|c|c|c}
\hline\hline
    & $\SO(\mathcal{N})_{\rm HC}$ & M10-PS & M11-PS \\
\hline\hline
$N_{\mathcal{N}}$ & $48 + 2^{\frac{\mathcal{N}-4}{2}}$ & $56$ & $50$ \\
$N_4$ & $3 (2 \mathcal{N}+1)$ & $63$ & $39$ \\
$N_L = N_R$ & $3 + 6 \mathcal{N} + 2^{\frac{\mathcal{N}-4}{2}}$ & $71$ & $41$ \\
 \hline 
\end{tabular} \end{center}
\caption{Multiplicity factors for the gauge couplings in the Pati-Salam UV completions.} \label{tab:NisPS}
\end{table}

In Fig.~\ref{fig:runningPS} we show the running of the Pati-Salam gauge couplings in the model M11-PS~\footnote{The usual matching applies: $$\alpha_4 = \alpha_3\,, \;\; \alpha_L = \alpha_2\,, \;\; \alpha_R = \frac{3 \alpha_1 \alpha_3}{3 \alpha_3 - 2 \alpha_1}\,.$$}: the first coupling to pass the safe threshold is $\alpha_R$, at an energy scale slightly lower than that for M11. In the numerical example, we fixed $\Lambda_{\rm Fl} = 10^{12}$~GeV, where the field content of M11-PS is added. Besides the difference in scales, the approach to the UV fixed point is similar, also showing the same fixed point for the two $\SU(2)$'s thanks to the left-right symmetry of the model.
For M10-PS, we find that the maximum allowed value for $\Lambda_{\rm Fl}$ is slightly above $10^{7}$~GeV, thus generating potential conflict with flavour bounds and also leaving too small space for the IR walking window. These results show that M11-PS is favoured.

We can now study the safety of the Yukawa couplings, which can be written in the Pati-Salam unified models as
\begin{eqnarray}
\mathcal{L}_{\rm Yuk, PS} &=& \left\{ \gamma_L^{ab}\ \Phi^{b,\ast} [\omega_L^a]_l [\Psi_L]_l + \gamma_R^{ab} \Phi^{b,\ast} [\omega_R]_r [\Psi_R]_r + \right. \nonumber \\
& \phantom{\frac{1}{2}}& \zeta_{Rr}^{ab}\ \Phi^a [\Xi_R^b]_l^c [\Psi_R]_r + \zeta_{Rl}^{ab}\ \Phi^a [\Xi_R^b]_r^c [\Psi_R]_l + \nonumber \\
& \phantom{\frac{1}{2}}& \zeta_{Lr}^{ab}\ \Phi^a [\Xi_L^b]_l^c [\Psi_L]_r + \zeta_{Ll}^{ab}\ \Phi^a [\Xi_L^b]_r^c [\Psi_L]_l + \nonumber \\
& \phantom{\frac{1}{2}}& \bar{\gamma}_L^{ab}\ \overline{\Phi}^{b,\ast} [\omega_L^a]_l [\Psi_L]_r^c + \bar{\gamma}_R^{ab} \overline{\Phi}^{b,\ast} [\omega_R]_r [\Psi_R]_l^c + \nonumber \\
& \phantom{\frac{1}{2}}& \bar{\zeta}_{Rr}^{ab}\ \overline{\Phi}^a [\Xi_R^b]_l^c [\Psi_R]_l^c + \bar{\zeta}_{Rl}^{ab}\ \overline{\Phi}^a [\Xi_R^b]_r^c [\Psi_R]_r^c + \nonumber \\
& \phantom{\frac{1}{2}}& \left. \bar{\zeta}_{Lr}^{ab}\ \overline{\Phi}^a [\Xi_L^b]_l^c [\Psi_L]_l^c + \bar{\zeta}_{Ll}^{ab}\ \overline{\Phi}^a [\Xi_L^b]_r^c [\Psi_L]_r^c \right\} \,. 
\label{eq:YukPS}
\end{eqnarray}
In absence of $\UU(1)$ couplings, the contribution of Yukawas and gauge couplings can be comparable. The beta function can be written as~\cite{Antipin:2018zdg,Vatani:2020}
\begin{multline}
(\beta_y)_{aij} = \frac{1}{32 \pi^2} \Big\{ (y_b \cdot y^{\dagger,b}\cdot y_a)_{ij} + (y_a \cdot y^{\dagger,b} \cdot y_b)_{ij} + \\
2 \mbox{Tr} [y_a \cdot y^{\dagger,b}] y_{bij}  \Big\} 
 -  \frac{3}{2} y_{aij} \sum_\alpha \frac{K_\alpha}{N_\alpha} H_0 (K_\alpha)  \times \\
 \left( \frac{C_2 (f_1) + C_2 (f_2)}{2} + \frac{C_2 (\Phi)}{12} K_\alpha \right)\,, 
\end{multline}
where $\alpha$ indicates the sum over the 4 gauge groups, and $C_2$ are the Casimirs of the \emph{irreps} of the two fermions and scalar under each gauge group. The function $H_0$ remains finite up to $K < 15/2$, thus the gauge contribution remains small up to the UV fixed points, reached for $K_\alpha = 3$ (where $H_0 (3) = 1/9$).
The beta function, after the gauge couplings have reached the fixed points, thus reads
\beq
\beta (y_k) = \frac{y_k}{32 \pi^2} \sum_{p} d_{kp} y_p^2 - C_{y_k} y_k\,,
\eeq
where $d_{kp}>0$ and order 1. Thus all Yukawas run to zero as long as $C_{y_k} > 0$ and $y_k$ is small enough at $\Lambda_{\rm Fl}$ that the beta functions are negative.
We find that the $C_{y_k}$ are the same for all the $\gamma$-type ($\bar{\gamma}$) and $\zeta$-type ($\bar{\zeta}$) Yukawas in Eq.~\eqref{eq:YukPS}, with
\beq \begin{array}{c}
C_\gamma = 0.046\,, \;\; C_\zeta = 0.066\,, \;\; \mbox{for M10-PS}\,;  \\
C_\gamma = 0.028\,, \;\; C_\zeta = 0.040\,, \;\; \mbox{for M11-PS}\,.
\end{array}
\eeq
As they are all positive and one order of magnitude larger than the factor $\frac{1}{32 \pi^2}$, the Yukawas are asymptotically free in both models with values of $\mathcal{O} (1)$ allowed at $\Lambda_{\rm Fl}$.

\section{Dark Matter phenomenology}
\label{DM}

\begin{figure*}[!hbt]
\centering
\includegraphics[width=0.9\textwidth]{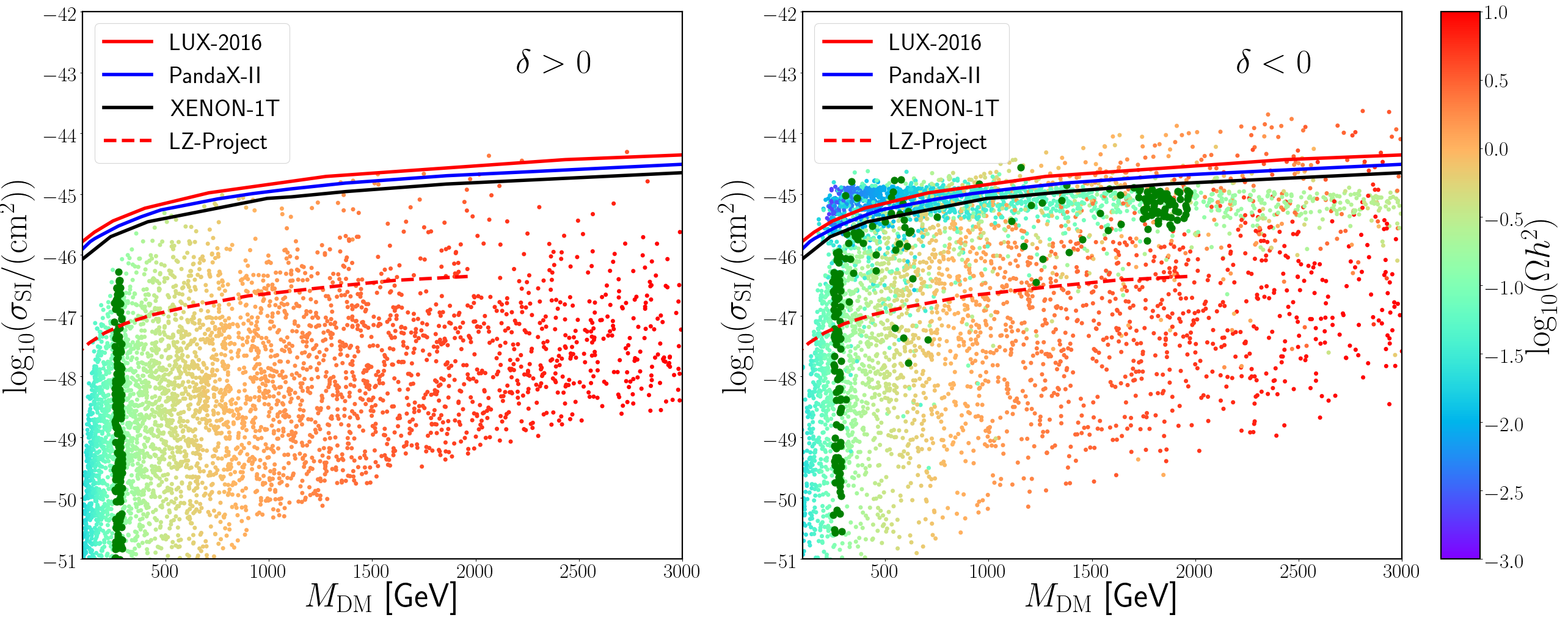}
\caption{Direct detection constraints~\cite{Cui:2017nnn,Akerib:2016vxi,Aprile:2018dbl} for $\delta > 0$ (left) and $\delta < 0$ (right).  The colour encodes the relic density for each parameter points, which is used to rescale the spin-independent cross section $\sigma_{\rm SI}$. The green curve saturates the measured relic density value, with points having warmer colour being excluded by over-density.}
\label{fig:DMDD}
\end{figure*}

The low energy physics of the two models, M10 and M11, can be described by the same effective field theory as they 
have the same global symmetries. The only distinction, besides the value of the low energy constants, can be traced in the properties
of the pNGBs associated to the global $\UU(1)$ symmetries broken by the condensates~\cite{Belyaev:2016ftv,Cacciapaglia:2019bqz}.
The Dark Matter sector is similar to that of the model in Ref.~\cite{Ma:2015gra}: the odd pNGBs consist of
a triplet of $\SU(2)_L$, an inert Higgs
doublet and charged and neutral singlets (forming a triplet of the custodial $\SU(2)_R$). However, the pNGB potential generated by the top 
interactions is very different, as here we use partial compositeness to generate the top mass, while in Ref.~\cite{Ma:2015gra} bilinear
Yukawa-like interactions are considered. Thus, mass spectra and couplings are different from those in  the model of Ref.~\cite{Ma:2015gra}. 
In this work we explicitly computed loops of the top and top partners after imposing the maximal symmetry~\cite{Csaki:2017cep} to keep 
the loops calculable and finite.

The nature of the lightest neutral stable scalar crucially depends on the masses of the preons: here we assume that the $\psi_U$
and $\psi_D$ have a common mass $m_R$ in order to preserve the global custodial $\SU(2)_R$, while the mass of $\psi_Q$ is $m_L$.
The crucial parameter is thus the mass difference 
$\delta \equiv (m_{L}-m_{R})/(m_{L}+m_{R})$. For $\delta < 0$, it mostly coincides with the $\SU(2)_L$
triplet, while for $\delta > 0$ it has maximal overlap with the singlets. For $\delta \sim 0$, maximal mixing with the doublet
is active. This mixing pattern determines the annihilation rates of the Dark Matter candidate, which is dominated by the final
states in two EW gauge bosons and two tops. The annihilation cross section is thus larger for $\delta < 0$, leading to larger
allowed Dark Matter masses.

To study a concrete example, we computed the top and gauge boson loop potential, considering top partners in the anti-symmetric of the unbroken $\SU(4)$, as embedded in the bi-fundamental (see~\ref{app:med} for more details). We constrain the parameter space by fixing the masses of the
top ($173$~GeV) and Higgs ($125$~GeV) at the minimum of the potential. We then scan the remaining parameter space and compute relic
abundance and spin-independent cross section off nuclei by using the {\tt micrOMEGAs}~\cite{Belanger:2013oya} package. For the misalignment
angle, $\sin \theta = \frac{v}{f}$, we probe values between $0.0003$ and $0.3$. Here, $v = 246$~GeV is the SM Higgs vacuum expectation value, $f \sim \frac{\Lambda_{\rm HC}}{4\pi}$ the decay constant of the composite Higgs, and the Dark Matter mass is proportional to $f$. In Fig.~\ref{fig:DMDD}
we show the results of our scan in the plane of the Dark Matter mass versus the cross section rescaled by the actual relic abundance.
In this way, all points can be compared to the Direct Detection exclusion, shown by the solid black, blue and red lines. Points
that saturate the relic abundance within 10$\sigma$ are highlighted in dark green. We see that the model can explain the Dark Matter abundance without
being excluded for $M_{\rm DM} \approx 260$~GeV, while larger values up to $1.5 \sim 2$~TeV are allowed for $\delta < 0$. Note also that the larger Dark Matter masses will be probed by future direct detection experiments.
We remark that those masses have reasonable values compared to the typical compositeness scale at the TeV.
We also performed a similar scan for top partners in the symmetric of the unbroken $\SU(4)$, finding that all points saturating
the relic density are excluded by direct detection.
Having demonstrated that a feasible Dark Matter candidate is present in the model, we leave a detailed study of the low energy
phenomenology of these models for future work.

\section{Conclusions and Outlook}
\label{sec:conclusion}

We have presented a new paradigm that allows to define composite Higgs models with partial compositeness
for the top quark up to arbitrarily high scales. For the first time, we can endow composite models with
predictivity power. Based on gauge-fermion underlying descriptions of the low energy physics, we use the need
for a large multiplicity of fermions, related to the large number of fermions and generations in the Standard Model,
to predict the presence of UV safe fixed points for the complete theory.

We apply this paradigm to models that also features a composite scalar Dark Matter candidate. We show that
the gauge couplings, which include the coupling of the confining group $\SO(10)_{\rm HC}$ (for M10) or $\SO(6)_{\rm HC}$ (for M11), can develop a UV
interacting fixed point while also allowing for an IR conformal window and a sufficient hierarchy between the
scale of flavour physics generation and the EW scale. Furthermore,  a
Dark Matter candidate is predicted in a consistent mass ballpark, which can also saturate the relic abundance while
evading direct detection bounds.

In the paradigm we propose, the four fermion interactions corresponding to partial compositeness
for the Standard Model fermions are generated by scalar mediators with a mass close to $\Lambda_{\rm Fl}$,
i.e. the scale where the theory approaches the UV fixed point. We showed that the Yukawa couplings run
to zero in the UV, thus not spoiling the safety of the model.
The well known instability on the scalar quartic couplings can be cured by embedding the model in a
Pati-Salam envelope above $\Lambda_{\rm Fl}$. We have shown that the two models, M10-PS and M11-PS,
also feature an interacting fixed point for the gauge couplings, and asymptotically free Yukawas, while M11-PS
based on $\SO(6)_{\rm HC}$ is preferred due to the higher flavour scale.
This also leaves open the possibility that the four-fermion interactions are generated by vector mediators, {\it \`a la}
Extended Technicolor. We leave the investigation of this point for further work.

The results presented in this work are stepping stones towards complete composite Higgs models, where the origin
of the standard fermion masses can be finally addressed. Some crucial ingredients, like the presence of an IR window
where large anomalous dimensions are generated, need input from lattice calculations, possible as a detailed underlying
model is on the table. The collider phenomenology of composite models is also affected, as non-minimal cosets are
the norm in this scenario, thus predicting additional charged and neutral light scalars that can be searched for at the LHC.

\section*{Acknowledgements}
Y.W. is supported by the Natural Sciences and Engineering Research Council of Canada (NSERC). T.M. is supported in part by project Y6Y2581B11 supported by 2016 National Postdoctoral Program for Innovative Talents.
G.C. and S.V. acknowledge partial support from the China-France LIA FCPPL and the Labex Lyon Institute of the Origins - LIO.

\appendix

\section{Conformal window} \label{app:confo}

To estimate if the model below $\Lambda_{\rm Fl}$ is inside the IR conformal window, we will utilise the Schwinger-Dyson rainbow approximation~\cite{Maskawa:1974vs,Fukuda:1976zb}, and the beta-function at two loops to estimate the fixed point. Following Ref. \cite{Sannino:2009za}, the beta functions read:
\beq
\beta_0 &=& \frac{11}{3} C_2(G) - \frac{4}{3} \sum_{i=\psi, \chi} T(r_i) n_i \,, \\
\beta_1 &=& \frac{34}{3} C_2^2 (G) - \frac{4}{3} \sum_{i=\psi, \chi} \left[5 C_2 (G) + 3 C_2 (r_i)  \right] T(r_i) n_i\,. \nonumber
\eeq
The IR fixed point, if existent, is characterised by the gauge coupling:
\beq
\alpha^\ast = - 4 \pi \frac{\beta_0}{\beta_1}\,,
\eeq
assuming that $\beta_1 < 0$ (and $\beta_0 >0$ as the theory is asymptotically free). In the rainbow approximation, we can calculate the value of the gauge coupling where the two specie condensates become critical: as the theory flows from an UV free point, the critical value is given by the smallest value of the two condensates:
\beq
\alpha_c = \mbox{min} \left\{ \frac{\pi}{3 C_2 (r_i)} \right\}\,.
\eeq
Equating $\alpha_c \equiv \alpha^\ast$ determines the lower edge of the conformal window. For the two models M10 and M11, fixing $n_\psi = 4$, we can find the range of $n_\chi$ leading to a theory inside the IR conformal window:
\beq
\SO(10) & \Rightarrow & 4 < n_\chi < 14\,, \\
\SO(6) & \Rightarrow & 6 < n_\chi < 9\,,
\eeq
where the upper edge is determined by the loss of asymptotic freedom. The models in Table~\ref{tab:M10} have $n_\chi = 8$ below $\Lambda_{\rm Fl}$, thus they are expected to be well inside the conformal window. 

As large anomalous dimensions are needed to enhance the top partial compositeness in particular, it may be needed to push the theory closer to the lower edge, where the couplings is stronger. In M10, for instance, one could push the mass of $\chi_d^3$ close to $\Lambda_{\rm Fl}$ in order to have $n_\chi = 5$ (while the bottom mass could be generated by the chimera baryons containing $\chi_u^3$). For M11, one could replace $\chi_l^3$ with an $\SU(2)_L$ singlet with hypercharge $-1$, thus leading to $n_\chi = 7$: this will marginally affect the running above $\Lambda_{\rm Fl}$, while only two neutrino masses can be generated (i.e., predicting one massless neutrino).

An alternative method to determine the conformal window, proposed in Ref.~\cite{Ryttov:2009yw}, is based on al all-order beta function conjecture, and it would lead to a lower edge for the conformal window, leading to $n_\chi > 3$ for both models.

\section{Resummation} \label{app:resum}

The functions appearing in the gauge coupling running are defined as~\cite{Vatani:2020}:
\beq
G_1 (K) &=&  \frac{3}{4} \int_0^K dx\ \tilde{F} (0,\frac{2}{3} c)\ \tilde{g} (\frac{1}{3} x)\,,   \nonumber \\
F_1 (K) &=&  \frac{3}{4} \int_0^K dx\ \tilde{F} (0,\frac{2}{3} c)\,, 
\eeq
where
\beq
\tilde{F} (0,y) &=& \frac{(1-y) (1-\frac{y}{3}) (1+\frac{y}{2}) \Gamma (4-y)}{3 \Gamma^2 (2-\frac{y}{2}) \Gamma (3-\frac{y}{2}) \Gamma (1+ \frac{y}{2})}\,, \nonumber \\
\tilde{g} (y) &=& \frac{20-43 y + 32 y^2 - 14 y^3 + 4 y^4}{4 (2y-1)(2y-3)(1-y^2)}\,. \nonumber
\eeq
We remark that the pole in $K=15/2$ comes from the $\Gamma (4-y)$--factor in $\tilde{F}$, which diverges for $y \to 5$, 
while the pole in $K=3$ for $G_1 (K)$ comes from the factor $(1-y^2)$ at the denominator of $\tilde{g}$. We remind the reader that 
$F_1 (K)$ corresponds to the resummation for abelian gauge groups~\cite{PalanquesMestre:1983zy}, and it encodes 2-loop diagrams
with fermion bubbles inserted in the gauge propagators. On the other hand, $G_1 (K)$ includes the contribution of 2-loop diagrams involving
gauge boson self couplings~\cite{Gracey:1996he,Holdom:2010qs}, with fermion bubble insertions. It is useful to connect our definitions with the function $H_1$ defined in Ref.~\cite{Antipin:2018zdg}:
\beq
H_1 = \frac{C_2 (G)}{C_2 (R_f)} \left( -\frac{11}{4} + G_1 \right) + F_1\,. 
\eeq

For the UV completions of the models M10 and M11, we find:
\beq
c_{\mathcal{N},\mathcal{N}} &=& C_2 ({\bf F}) = \frac{\mathcal{N}-1}{2}\,, \\
c_{\mathcal{N},3} &=& 24 \frac{T({\bf F})}{N_\mathcal{N}}\,, \\
c_{\mathcal{N},2}  &=& \frac{3 T({\bf F})}{2 N_{\mathcal{N}}} \left( 3 + \frac{T ({\bf Sp})}{T({\bf F})} \right)\,, \\
c_{\mathcal{N},1}  &=& \frac{T({\bf F})}{N_{\mathcal{N}}} \left( \frac{13}{2} + \frac{T ({\bf Sp})}{2 T({\bf F})} \right)\,, 
\eeq
\beq
c_{3, \mathcal{N}} &=& 3 \frac{T({\bf F}) d({\bf G})}{N_3}\,, \\
c_{3,3} &=& \frac{4}{3}\,, \\
c_{3,2} &=& \frac{9}{8 N_3}\,, \\
c_{3,1} &=& \frac{1}{N_3} \left( \frac{5}{6} d ({\bf F}) + \frac{11}{24} \right)\,, 
\eeq
\beq
c_{2,\mathcal{N}} &=& \frac{T({\bf F}) d ({\bf G})}{N_2} \left(\frac{3}{2} + \frac{T({\bf Sp})}{2 T({\bf F})} \right)\,, \\
c_{2,3} &=& \frac{3}{N_2}\,, \\
c_{2,2} &=& \frac{3}{4}\,, \\
c_{2,1} &=& \frac{1}{N_2} \left( \frac{3}{8} d ({\bf F}) + \frac{1}{4} \right)\,, 
\eeq
\beq
c_{1,\mathcal{N}} &=& \frac{T({\bf F}) d({\bf G})}{N_1} \left( \frac{13}{2} + \frac{T({\bf Sp})}{2 T({\bf F})} \right)\,, \\
c_{1,3} &=& \frac{1}{N_1} \left( \frac{20}{3} d({\bf F}) + \frac{11}{3} \right)\,, \\
c_{1,2} &=& \frac{1}{N_1} \left( \frac{9}{8} d({\bf F}) + \frac{3}{4} \right)\,, \\
c_{1,1} &=& \frac{1}{N_1} \left( \frac{163}{72} d({\bf F}) + \frac{1}{8} d({\bf Sp}) + \frac{95}{36} \right)\,.
\eeq
The group theory factors appearing in the above expressions refer to the $\SO(\mathcal{N})$ \emph{irreps}, and are equal to
$$
\begin{array}{ccc}
d({\bf G}) = \frac{\mathcal{N} (\mathcal{N}-1)}{2}\,, & d({\bf F}) = \mathcal{N}\,, & d({\bf Sp}) = 2^\frac{\mathcal{N}-2}{2}\,, \\
T({\bf G}) = \mathcal{N}-2\,, & T({\bf F}) = 1\,, & T({\bf Sp}) = 2^\frac{\mathcal{N}-8}{2}\,, \\
C_2({\bf G}) = \mathcal{N}-2\,, & C_2({\bf F}) = \frac{\mathcal{N}-1}{2}\,, & C_2({\bf Sp}) = 2^\frac{\mathcal{N} (\mathcal{N}-1)}{16}\,.
\end{array}
$$
Numerically, for M10 we find: 
\beq
c_{i,j}^{M10} = \begin{pmatrix}
4.5 & 0.75 & 0.234 & 0.234 \\
4.09 & 1.33 & 0.068 & 0.266 \\
4.33 & 0.115 & 0.75 & 0.154 \\
4.33 & 0.90 & 0.154 & 0.350
\end{pmatrix}\,;
\eeq
while for M11:
\beq
c_{i,j}^{M11} = \begin{pmatrix}
2.5 & 0.923 & 0.202 & 0.260 \\
2.14 & 1.33 & 0.107 & 0.260 \\
1.88 & 0.214 & 0.75 & 0.179 \\
2.20 & 0.949 & 0.163 & 0.364
\end{pmatrix}\,;
\eeq
where $i,j = \mathcal{N},3,2,1$.


\section{Top partners} \label{app:med}

For the partial compositeness, based on the UV completions in Table~\ref{tab:M10}, we are considering the case where composite top partners transform in the antisymmetric representations $6$ and $\bar{6}$ under unbroken subgroup $\SU(4)$ of the global symmetry $\SU(4)_L\times \SU(4)_R$. In order to preserve the Dark parity in the theory, we shall include both representations in a symmetric way. 
To write the proper mixing terms, the elementary top quark fields need to be embedded in the above representations, by way of the following spurions:
\begin{align}
\psi_{q_L}^{u,6} &= \frac{1}{\sqrt{2}}\left(\begin{array}{cccc}
0 & 0 & 0 & t_L \\
0 & 0 & 0 & b_L \\
0 & 0 & 0 & 0 \\
-t_L & -b_L & 0 & 0
\end{array}\right),\nonumber\\
\psi_{q_L}^{d,6} &= \frac{1}{\sqrt{2}}\left(\begin{array}{cccc}
0 & 0 & t_L & 0 \\
0 & 0 & b_L & 0 \\
-t_L & -b_L & 0 & 0 \\
0 & 0 & 0 & 0
\end{array}\right), \nonumber \\
\psi_{q_L}^{u,\bar{6}} &= \frac{1}{\sqrt{2}}\left(\begin{array}{cccc}
0 & 0 & -b_L & 0 \\
0 & 0 & t_L & 0 \\
b_L & -t_L & 0 & 0 \\
0 & 0 & 0 & 0
\end{array}\right),\nonumber \\
\psi_{q_L}^{d,\bar{6}} &= \frac{1}{\sqrt{2}}\left(\begin{array}{cccc}
0 & 0 & 0 & b_L \\
0 & 0 & 0 & -t_L \\
0 & 0 & 0 & 0 \\
-b_L & t_L & 0 & 0
\end{array}\right) \nonumber \\
\psi_{t_R}^{6(\bar{6})} &=\frac{1}{\sqrt{2}}\left(\begin{array}{cccc}
0 & t_R & 0 & 0 \\
-t_R & 0 & 0 & 0 \\
0 & 0 & 0 & 0 \\
0 & 0 & 0 & 0
\end{array}\right), \nonumber \\ 
\psi_{b_R}^{6(\bar{6})} &=\frac{1}{\sqrt{2}}\left(\begin{array}{cccc}
0 & b_R & 0 & 0 \\
-b_R & 0 & 0 & 0 \\
0 & 0 & 0 & 0 \\
0 & 0 & 0 & 0
\end{array}\right)
\end{align}
The transformation properties of these spurions are
\begin{align}
\psi_{q_L}^{u/d,6} &\to L \psi_{q_L}^{u/d,6} L^T, \nonumber\\
\psi_{q_L}^{u/d,\bar{6}} &\to R^*\psi_{q_L}^{u/d,\bar{6}} R^\dagger, \nonumber \\
\psi_{t_R/b_R}^{6} &\to L \psi_{t_R/b_R}^{6} L^T, \nonumber\\
\psi_{t_R/b_R}^{\bar{6}} &\to R^* \psi_{t_R/b_R}^{\bar{6}} R^\dagger,
\end{align}
where $L$ ($R$) is an element of the global symmetry 
$\SU(4)_L$ ($\SU(4)_R$). 

At low energies, the partial compositeness Lagrangian can be written as
\begin{align}
        \label{equ:PCLag}
        \mathcal{L} &= \epsilon_L^{u,6}f{\rm Tr}[\bar{\psi}_{q_L}^{u,6}U\mathcal{B}_{q_L}^{u,6}U^T] + \epsilon_L^{u,\bar{6}}f{\rm Tr}[\bar{\psi}_{q_L}^{u,\bar{6}}U^T\mathcal{B}_{q_L}^{u,\bar{6}}U] \nonumber \\
        &+ \epsilon_L^{d,6}f{\rm Tr}[\bar{\psi}_{q_L}^{d,6}U\mathcal{B}_{q_L}^{d,6}U^T] + \epsilon_L^{d,\bar{6}}f{\rm Tr}[\bar{\psi}_{q_L}^{d,\bar{6}}U^T\mathcal{B}_{q_L}^{d,\bar{6}}U] \nonumber \\
        &+ \epsilon_R^{u,6}f{\rm Tr}[\bar{\psi}_{t_R}^{6}U\mathcal{B}_{t_R}^{6}U^T]+\epsilon_R^{u,\bar{6}}f{\rm Tr}[\bar{\psi}_{t_R}^{\bar{6}}U^T\mathcal{B}_{t_R}^{\bar{6}}U] \nonumber \\
        &+ \epsilon_R^{d,6}f{\rm Tr}[\bar{\psi}_{b_R}^{6}U\mathcal{B}_{b_R}^{6}U^T]+\epsilon_R^{d,\bar{6}}f{\rm Tr}[\bar{\psi}_{b_R}^{\bar{6}}U^T\mathcal{B}_{b_R}^{\bar{6}}U] \nonumber \\
        &+ M_{6\bar{6}}^{u}{\rm Tr}[\bar{\mathcal{B}}_{q_L}^{u,6}\mathcal{B}_{t_R}^{\bar{6} \prime}] + M_{\bar{6}6}^{u}{\rm Tr}[\bar{\mathcal{B}}_{q_L}^{u,\bar{6}}\mathcal{B}_{t_R}^{6 \prime}] \nonumber \\
        &+ M_{6\bar{6}}^{d}{\rm Tr}[\bar{\mathcal{B}}_{q_L}^{d,6}\mathcal{B}_{b_R}^{\bar{6} \prime}] + M_{\bar{6}6}^{d}{\rm Tr}[\bar{\mathcal{B}}_{q_L}^{d,\bar{6}}\mathcal{B}_{b_R}^{6 \prime}] + h.c.
\end{align}
where the $\mathcal{B}$'s are the corresponding top partners, $\mathcal{B}^{ \prime}_{ij} =\epsilon_{ij kl} \mathcal{B}_{kl}$ and 
$U$ is the usual non-linear sigma field, which transforms under the global symmetry as 
\begin{align}
U &\to L U h^\dagger\,, \quad U \to h U R^\dagger\,,
\end{align}
where $h$ is an element of the unbroken group $\SU(4)$. In order to preserve the Dark parity, the following conditions must be satisfied:
\begin{align}
\label{equ:DMCondition}
\epsilon_L^{u,6} = -\epsilon_L^{u,\bar{6}} \equiv \epsilon_L^u,&\quad\epsilon_L^{d,6} =- \epsilon_L^{d,\bar{6}} \equiv \epsilon_L^d \nonumber\\
\epsilon_R^{u,6} = \epsilon_R^{u,\bar{6}} \equiv \epsilon_R^u,&\quad\epsilon_R^{d,6} = \epsilon_R^{d,\bar{6}} \equiv \epsilon_R^d \nonumber\\
 M_{6\bar{6}}^u = M_{\bar{6}6}^u \equiv M_\Delta^u, &\quad M_{6\bar{6}}^d = M_{\bar{6}6}^d \equiv M_\Delta^d.
\end{align}

It is easy to verify that the underlying Lagrangian is invariant under the DM parity with following transformations:
\begin{align}
        \psi_{X}^{6} &\leftrightarrow -P_B ( \psi_{X}^{\bar{6}} ) P_B^\dagger =\psi_{X}^{6}\,, \nonumber \\
        \mathcal{B}_{X}^{6} &\leftrightarrow P_B \mathcal{B}_{X}^{\bar{6}} P_B^\dagger\,, \quad   U \to  P_B U^T P_B^\dagger\,;
\end{align}        
where $P_B$ is the Dark parity transformation defined in~\cite{Ma:2015gra}:
$$
P_B = \left( \begin{array}{cc}
\sigma_2 & 0 \\ 0 & -\sigma_2 \end{array}\right).
$$ 

The scalar potential has a similar for to that used in~\cite{Ma:2017vzm,Ma:2015gra}, with the fermion-induced potential replaced by the one induced bu loops of the above partial compositeness Lagrangian. The full potential has been coded into {\tt FeynRules}~\cite{Alloul:2013bka}, and then matched on to {\tt micrOMEGAs} \cite{Barducci:2016pcb}. The latter is used to calculate the Dark Matter relic density as well as the scattering cross-sections.

\bibliographystyle{JHEP-2-2}

\bibliography{FundHDM.bib}

\end{document}